\documentclass[aps,prd,twocolumn,showpacs]{revtex4} 

\usepackage{graphicx}

\begin{document}

\title{Lepton Number-Driven Sterile Neutrino Production in the Early Universe}
\author{Chad T. Kishimoto \& George M. Fuller}
\affiliation{Department of Physics, University of California, San Diego, La Jolla, California
92093-0319, USA}

\date{\today}

\begin{abstract}
We examine medium-enhanced, neutrino scattering-induced decoherent production of dark matter candidate sterile neutrinos in the early universe.  In cases with a significant net lepton number we find two resonances, where the effective in-medium mixing angles are large.  We calculate the lepton number depletion-driven evolution of these resonances.  We describe the dependence of this evolution on lepton numbers, sterile neutrino rest mass, and the active-sterile vacuum mixing angle.  We find that this resonance evolution can result in relic sterile neutrino energy spectra with a generic form which is sharply peaked in energy.   We compare our complete quantum kinetic equation treatment with the widely-used quantum Zeno ansatz.
\end{abstract}
\maketitle

\section{Introduction}
\label{sec:introduction}

In this paper we examine the physics which determines the relic densities and fossil energy spectra of neutrinos in the very early universe.  The existence of new  kinds of electroweak singlet ``sterile'' neutrinos, $\nu_s$, with subweak interactions remains a controversial possibility.  The experimental establishment of nonzero neutrino rest masses fuels legitimate speculation on this issue.  

Moreover, this issue could be important because sterile neutrinos with rest masses $m_s \sim 1 \,{\rm keV}$ are interesting dark matter candidates.  The minimalist model for this dark matter scenario is where sterile neutrinos are coupled to active species via mass terms that give vacuum mixing.  In turn, the simplest model along these lines is where active neutrino scattering-induced decoherence gives rise to a relic density of sterile neutrinos \cite{dw94}.  

These simple models, however, are challenged.  Vacuum active-sterile neutrino mixing, which is necessary to produce the sterile neutrinos from an initially purely active neutrino population, also enables a non-GIM suppressed radiative decay channel ($\nu_s \rightarrow \nu_{e, \mu, \tau} + \gamma$) for sterile neutrinos.  As a consequence, the best constraints and probes of this sector of particle physics come from the modern x-ray observatories \cite{aft, afp, dh02, ak06, bnos06, bnos06b, bnost06, pcss07,ybw07, wbyw06}.  These constraints arguably eliminate the simplest models for decoherence-produced sterile neutrino dark matter.

However, there remain processes for sterile neutrino production which can produce the correct relic density for dark matter yet evade all current bounds.  These viable models include, for example, sterile neutrino production associated with inflation \cite{st06}, Higgs decay \cite{kusenko06, pk07, petraki08}, and lepton number-driven medium enhancement \cite{sf99, afp}.

The concept of lepton number is a slippery one if sterile neutrinos exist and mix in vacuum with active species.  In essence, lepton number can be created \cite{ftv96, shi96} or destroyed \cite{abfw, kfs06} by active-sterile neutrino interconversion.  The current observational bounds \cite{kssw01, abb02, wong02, dhpprs02, cuo04, sr05, ss08, abfw} on the electron, muon, and tau lepton numbers of the universe are relatively poor, at least on the scale of these quantities required to affect active-sterile neutrino conversion.  Therefore, it may be important to examine in detail how lepton number influences scattering-induced decoherent production of sterile neutrinos in the early universe.   This is the objective of this paper.

In what follows we investigate how the coupled interplay of sterile neutrino production, lepton number depletion, and the expansion of the universe determine the relic sterile neutrino densities and energy spectra.  In Sec.\ \ref{sec:resonances} we describe in-medium active-sterile resonances and how these influence decoherent sterile neutrino production.  In this section we also discuss the relationship between the quantum kinetic equations and the quantum Zeno approximation.  In Sec.\ \ref{sec:calculations} we discuss the approach we employ to solve for lepton number and $\nu_{\rm s}$ spectral evolution.  In Sec.\ \ref{sec:results} we discuss results.  Conclusions are given in Sec.\ \ref{sec:conclusions}.

\section{Sterile Neutrino Production in the Early Universe}
\label{sec:resonances}

There is a long history to the general problem of the production of sterile neutrinos $\nu_s$ from an initial population of purely active neutrinos $\nu_\alpha$ ($\alpha = e, \mu, \tau$) \cite{afp, dw94, sf99, ftv96, fv95, fv97, df02}.  In low density environments in supernovae and the early universe, the dominant $\nu_s$ production channel will be the coherent Mikheyev-Smirnov-Wolfenstein (MSW) process \cite{MS, W}.  However, neutrino propagation may not be coherent in the higher density regions of core collapse supernovae and the pre-neutrino-decoupling epoch in the early universe.  In these environments, scattering-induced decoherence is the principal way in which sterile neutrinos are produced.  Here we concentrate on the decoherent production channel since this will dominate the production of dark matter candidate sterile neutrinos.

\subsection{The Quantum Kinetic Equations}
\label{sec:qke}
The evolution of a system of active and sterile neutrinos can be formulated in terms of the density operator for a neutrino with given scaled momentum $\epsilon \equiv p_\nu / T$, with $p_\nu$ the neutrino momentum and $T$ the temperature in the early universe.  (Here we use natural units in which $\hbar = c = k_{\rm B} = 1$.)  The density operator for scaled momentum $\epsilon$ is
\begin{equation}
\rho (\epsilon, t) = \sum_{i, j} \rho_{ij} ( \epsilon, t ) \vert \nu_i \rangle \langle \nu_j \vert ,
\end{equation}
where the summation is performed over neutrino flavor states, {\it e.g.}, $\vert \nu_i \rangle$ with $i = e, \mu, \tau, s$ and $\rho_{ij} (\epsilon, t)$ are the corresponding density operator matrix elements which are momentum and time dependent.  Here we take the temperature $T (t)$ to be a function of the Friedman-Lema\^itre-Robertson-Walker time coordinate $t$, defined by the solution of the zero-curvature Friedman equation
\begin{equation}
\left( \frac{\dot{a}}{a} \right)^2 = \frac{8 \pi \rho (t)}{3 m_{\rm PL}^2},
\end{equation}
where $a(t)$ is the scale factor, $\rho(t)$ is the total energy density of the universe, and $m_{\rm PL}$ is the Planck mass.  

For illustrative purposes, we consider $2 \times 2$ vacuum mixing between an active and sterile neutrino with a one parameter (vacuum mixing angle $\theta$) unitary transformation between the weak eigenstates $\vert \nu_\alpha \rangle$ ($\alpha = e, \mu, \tau$), $\vert \nu_s \rangle$ and the energy-mass eigenstates $\vert \nu_1 \rangle$, $\vert \nu_2 \rangle$:
\begin{eqnarray}
\vert \nu_\alpha \rangle & = & \cos \theta \vert \nu_1 \rangle + \sin \theta \vert \nu_2 \rangle ; \nonumber \\
\vert \nu_s \rangle & = & - \sin \theta \vert \nu_1 \rangle + \cos \theta \vert \nu_2 \rangle .
\end{eqnarray}
In this $2 \times 2$ formalism, it is convenient to decompose the density operator as
\begin{equation}
\rho ( \epsilon, t ) = \frac{1}{2} P_0 ( \epsilon, t ) \left[ 1 + \mathbf{P} ( \epsilon, t ) \cdot \mbox{\boldmath$\sigma$} \right] ,
\label{eq:rhodecomposition}
\end{equation}
where $\mbox{\boldmath{$\sigma$}}$ is the Pauli spin operator, $P_0 (\epsilon, t)$ is a normalization factor proportional to the total number of neutrinos (active and sterile) with scaled momentum $\epsilon$, and $\mathbf{P} (\epsilon, t)$ acts as a polarization vector in weak isospin space.

The diagonal matrix elements of the density operator,
\begin{eqnarray}
f_\alpha ( \epsilon, t ) \equiv \rho_{\alpha \alpha} (\epsilon, t) & = & \frac{1}{2} P_0 ( \epsilon, t ) \left[ 1 + P_z ( \epsilon, t ) \right]  \\
f_s ( \epsilon, t ) \equiv \rho_{s s} (\epsilon, t) & = & \frac{1}{2} P_0 ( \epsilon, t ) \left[ 1 - P_z ( \epsilon, t ) \right] ,
\end{eqnarray}
are proportional to the number density distributions of the neutrino species.  The density operators are normalized so that the neutrino number densities are given as products of the above matrix elements and the zero chemical potential Fermi-Dirac distribution:
\begin{eqnarray}
n_\alpha ( \epsilon, t )  & = & f_\alpha (\epsilon, t) \frac{T^3 (t)}{2 \pi^2} \frac{\epsilon^2}{e^{\epsilon} + 1} ; \\
n_s ( \epsilon, t ) & =  & f_s (\epsilon, t) \frac{T^3 (t)}{2 \pi^2} \frac{\epsilon^2}{e^{\epsilon} + 1} .
\end{eqnarray}

The quantum kinetic equations can be derived \cite{mt94, sr93} from the time evolution of the density operator.  In a homogeneous and isotropic universe, these equations can be given in terms of the time evolution of the functions $P_0 (\epsilon, t)$ and $\mathbf{P} ( \epsilon, t )$ as defined in Eq.\ (\ref{eq:rhodecomposition}),
\begin{eqnarray}
\frac{\partial}{\partial t} \mathbf{P} ( \epsilon, t ) & = & \mathbf{V} ( \epsilon, t ) \times \mathbf{P} ( \epsilon, t ) \nonumber \\ & &+  \left[ 1 - P_z ( \epsilon, t ) \right] \left[ \frac{\partial}{\partial t} \ln P_0 ( \epsilon, t ) \right] \hat{z} \nonumber \\
 & & - \left[ D( \epsilon, t ) + \frac{\partial}{\partial t} \ln P_0 ( \epsilon, t ) \right] \mathbf{P}_{\perp} ( \epsilon, t ) 
\label{eq:qkePvec} \\
\frac{\partial}{\partial t} P_0 ( \epsilon, t ) & = & R ( \epsilon, t ) ,
\label{eq:qkeP0}
\end{eqnarray}
where $\mathbf{P}_\perp = P_x \hat{x} + P_y \hat{y}$, and the functions $\mathbf{V}, D,$ and $R$ are defined below.

The vector $\mathbf{V} ( \epsilon, t )$ corresponds to the coherent quantum mechanical evolution of the neutrino states in the early universe, and is given by
\begin{equation}
\mathbf{V} ( \epsilon, t ) = \frac{\delta m^2}{2 \epsilon T(t)} \left( \sin 2 \theta \,\hat{x} - \cos 2 \theta \,\hat{z} \right) + V_\alpha (\epsilon, t) \hat{z} ,
\end{equation}
where $\delta m^2 = m_2^2 - m_1^2$ is the difference of the squares of the vacuum neutrino mass eigenvalues.  The first term represents vacuum neutrino oscillations, while the second term introduces matter effects through the forward scattering potential \cite{nr88, fmws87}
\begin{equation}
V_\alpha (\epsilon, t) = \frac{2 \sqrt{2} \zeta (3)}{\pi^2} G_F \mathcal{L}_\alpha ( t ) T^3 (t) - r_\alpha G_F^2 \epsilon \,T^5 (t) ,
\label{eq:fwdscatteringpot}
\end{equation}
where $\zeta (3) \approx 1.20206$, $G_F$ is the Fermi constant, and $r_\alpha$ is a dimensionless coefficient which depends on the number of charged lepton degrees of freedom and can be found in Refs.\ \cite{abfw, afp}.  The potential lepton number is
\begin{equation}
\mathcal{L}_\alpha (t) \equiv 2 L_\alpha (t) + \sum_{\beta \neq \alpha} L_\beta (t) .
\end{equation}
The individual lepton numbers are
\begin{equation}
L_\alpha (t) = \frac{n_{\nu_\alpha} (t) - n_{\bar\nu_\alpha} (t)}{n_\gamma (t)},
\end{equation}
where $n_{\nu_\alpha}$, $n_{\bar\nu_\alpha}$, and $n_\gamma$ are the neutrino, antineutrino and photon number densities, respectively.  Here we neglect contributions to the forward scattering potential from neutrino-baryon and neutrino-charged-lepton interactions since we will consider lepton numbers much larger than the baryon to photon ratio \cite{afp, abfw}.

The decoherence function $D(\epsilon, t)$ and the repopulation function $R(\epsilon, t)$ can be simplified by assuming thermal equilibrium of the background plasma \cite{bvw99}.  The decoherence function corresponds to the loss of coherence resulting from collisions with particles in the early universe, and is proportional to the total scattering rate $\Gamma_\alpha ( \epsilon, t )$ of neutrinos $\nu_\alpha$,
\begin{equation}
D (\epsilon, t) = \frac{1}{2} \Gamma_\alpha ( \epsilon, t ) .
\end{equation}

Using the assumption of thermal equilibrium, the total scattering rate can be written as
\begin{equation}
\Gamma_\alpha ( \epsilon, t ) \approx y_\alpha (t) G_F^2 \epsilon \,T^5 (t) ,
\label{eq:totscatteringrate}
\end{equation}
where we have neglected corrections of order the lepton number $L_\alpha$.  The numerical coefficient $y_\alpha (t)$ primarily depends on the number of relativistic particles with weak charge that are populated in the thermal seas of the early universe at epoch $t$.  For example, at temperatures \mbox{$1 \,{\rm MeV} \lesssim T \lesssim 20 \,{\rm MeV}$} the total scattering rate mostly stems from interactions with other neutrinos and $e^\pm$ pairs, so that  $y_e \approx 1.27$ and $y_{\mu, \tau} \approx 0.92$ \cite{afp}.  However, at higher temperatures there can be appreciable populations of other charged leptons and quarks which will increase the total scattering rate.

The repopulation function $R (\epsilon, t)$ dictates the evolution of $P_0 (\epsilon, t)$.  Since $P_0$ is proportional to the total number of neutrinos with scaled momentum $\epsilon$, the repopulation function corresponds to scattering into and out of neutrino states ($\nu_\alpha$ and $\nu_s$) with scaled momentum $\epsilon$.  Assuming that each populated species, with the exception of $\nu_\alpha$ and $\nu_s$, has a thermal spectrum, the repopulation function can be written as
\begin{equation}
R ( \epsilon, t ) = \Gamma_\alpha ( \epsilon, t ) \left[ \frac{e^\epsilon + 1}{e^{\epsilon - \eta_\alpha (t)} + 1} - f_\alpha ( \epsilon, t ) \right] ,
\end{equation}
where $\eta_\alpha$ is the degeneracy parameter (ratio of the chemical potential to the temperature) associated with the Fermi-Dirac spectra of $\nu_\alpha$ and $\bar\nu_\alpha$ with lepton number $L_\alpha$,
\begin{equation}
L_\alpha = \frac{\pi^2}{12 \zeta(3)} \left( \eta_\alpha + \frac{1}{\pi^2} \eta_\alpha^3 \right) .
\end{equation}
This form of the repopulation function is valid for temperatures above the neutrino decoupling temperature, where the active neutrinos are able to efficiently exchange energy and momentum with the plasma of the early universe.

The effect of the repopulation function is to drive the distribution of the active neutrino toward a Fermi-Dirac spectrum consistent with a lepton number $L_\alpha$,
\begin{equation}
n_\alpha ( \epsilon, t ) \propto f_\alpha (\epsilon, t) \frac{\epsilon^2}{e^\epsilon + 1} \rightarrow \frac{\epsilon^2}{e^{\epsilon - \eta_\alpha (t)} + 1} .
\end{equation}

\subsection{The Quantum Zeno Ansatz}
The quantum kinetic equations, even in their simplified form with the assumptions of homogeneity and isotropy of the early universe and thermal equilibrium in the background plasma, comprise a system of coupled nonlinear integro-partial differential equations that are difficult to solve.  Past works have employed the quantum Zeno approximation to circumvent these difficulties \cite{afp, fv97}.  We write the quantum Zeno ansatz as a Boltzmann-like kinetic equation,
\begin{widetext}
\begin{equation}
\frac{\partial}{\partial t} f_s ( \epsilon, t ) + \left( \frac{d \epsilon}{d t} \right) \frac{\partial}{\partial \epsilon} f_s (\epsilon, t) \approx \frac{1}{4} \Gamma_\alpha (\epsilon, t) \sin^2 2 \theta_M ( \epsilon, t )\left[ 1 + \left( \frac{\Gamma_\alpha ( \epsilon, t ) \ell_m ( \epsilon, t )}{2} \right)^2 \right]^{-1} \left[ f_\alpha ( \epsilon, t ) - f_s ( \epsilon, t ) \right] ,
\label{eq:quantumzenoansatz}
\end{equation}
where the effective matter mixing angle $\theta_M$ is defined by
\begin{equation}
\sin^2 2 \theta_M ( \epsilon, t ) = \frac{V_x^2 ( \epsilon, t )}{\vert \mathbf{V} ( \epsilon, t ) \vert^2} = \sin^2 2 \theta \left[ \sin^2 2 \theta + \left( \cos 2 \theta - \frac{2 \epsilon T(t)}{\delta m^2} V_\alpha ( \epsilon, t) \right)^2 \right]^{-1}  ,
\end{equation}
and the neutrino oscillation length is
\begin{equation}
\ell_m ( \epsilon, t ) =  \vert \mathbf{V} ( \epsilon, t ) \vert^{-1} = \left[ \left( \frac{\delta m^2}{2 \epsilon T(t)} \sin 2 \theta \right)^2 + \left( \frac{\delta m^2}{2 \epsilon T(t)} \cos 2 \theta - V_\alpha (\epsilon, t) \right)^2 \right]^{-1/2} .
\end{equation}
\end{widetext}
The $d \epsilon / d t$ term takes into account the change in the scaled momentum of a neutrino propagating along its world line through the expanding early universe.  Using the conservation of comoving entropy and assuming radiation domination, this is
\begin{equation}
\frac{d \epsilon}{d t} = \frac{\epsilon}{3 g} \frac{d g}{d t},
\label{eq:varepsilon}
\end{equation}
where $g$ is the effective statistical weight in relativistic particles and is calculated with a weighted sum of the bosonic ($g_b$) and fermionic ($g_f$) degrees of freedom,
\begin{equation}
g = \sum_{\rm bosons} g_b + \frac{7}{8} \sum_{\rm fermions} g_f .
\end{equation}
For example, at temperatures \mbox{$1 \,{\rm MeV} \lesssim T \lesssim 20 \,{\rm MeV}$} the plasma of the early universe consists of photons ($g_b = 2$), $e^\pm$ pairs ($g_f = 4$), and neutrinos and antineutrinos of all three active flavors ($g_f = 6$), so $g \approx 10.75$, where we neglect corrections of order the lepton numbers.

Using the quantum Zeno ansatz instead of the full quantum kinetic equations eases computational demands by allowing calculations to be performed with distribution functions (amplitude squared) instead of with quantum amplitudes.  The former approach avoids the computational pitfalls of rapidly varying complex phases that plague the latter.  In Sec.\ \ref{sec:results}, we will compare the sterile neutrino production rates calculated by the quantum kinetic equations to those implied by the quantum Zeno ansatz.

Inspection of the quantum Zeno ansatz, Eq.\ (\ref{eq:quantumzenoansatz}), suggests that maximal sterile neutrino production should occur when $\sin^2 2 \theta_M = 1$.  This corresponds to the resonance condition in the coherent MSW process, 
\begin{equation}
\delta m^2 \cos 2 \theta = \frac{4 \sqrt{2} \zeta(3)}{\pi^2} G_F \mathcal{L} \epsilon_{\rm res} T^4 - 2 r_\alpha G_F^2 \epsilon^2_{\rm res} T^6 .
\label{eq:mswresonanceeqn}
\end{equation}
The solutions to Eq.\ (\ref{eq:mswresonanceeqn}) are the resonant scaled momenta
\begin{equation}
\epsilon_{\rm res} (t) = \frac{\sqrt{2} \zeta(3)}{\pi^2 r_\alpha G_F} \frac{\mathcal{L}(t)}{T^2(t)} \left( 1 \pm \sqrt{1 - \frac{\pi^4 r_\alpha \delta m^2 \cos 2 \theta}{4 \zeta^2 (3) \mathcal{L}^2 (t) T^2(t)}} \right) .
\label{eq:mswresonancesoln}
\end{equation}
Therefore, resonances occur when the following condition is met:
\begin{equation}
\left\vert \mathcal{L} (t) \right\vert T(t) \geq \frac{\pi^2}{2 \zeta(3)} \sqrt{r_\alpha \delta m^2 \cos 2 \theta}.
\label{eq:resonancecondition}
\end{equation} 

In Sec.\ \ref{sec:results} we will discuss the results of our numerical calculations and the effects of these MSW-like resonances and this resonance condition.  The production of sterile neutrinos from an initial active neutrino distribution reduces the potential lepton number $\mathcal{L}$.  Furthermore, the expansion of the universe results in a decrease of the temperature $T$.  These two trends together ensure that the resonance condition, Eq.\ (\ref{eq:resonancecondition}), will be violated at some point.  However, the enhancement in sterile neutrino production associated with this MSW resonance is suppressed by the quantum Zeno effect.  

The quantum Zeno effect is the result of scattering-induced decoherence interrupting the accumulation of quantum phase, suppressing quantum transitions between two discrete states \cite{zeno}.  In the quantum Zeno ansatz, this effect is represented by the multiplicative factor,
\begin{displaymath}
\left[ 1 + \left( \frac{\Gamma_\alpha \ell_m}{2} \right)^2 \right]^{-1} .
\end{displaymath}
The ratio of the neutrino oscillation length, $\ell_m$, to the mean scattering length, $\Gamma_\alpha^{-1}$, determines the level of suppression stemming from the quantum Zeno effect.  

If the neutrino oscillation length is much larger than the scattering length, $\ell_m \gg \Gamma_\alpha^{-1}$, quantum coherence is lost before significant quantum phase can be accumulated, suppressing the transition.  This is the case at the MSW resonances because at resonance the neutrino oscillation length is maximal.  The suppression of sterile neutrino production at the MSW resonances means that maximal sterile neutrino production occurs at a time that does not correspond to an MSW resonance.  However, since we will be considering small values of $\sin^2 2 \theta$, the MSW resonances have small widths and as a result we nevertheless have maximal sterile neutrino production near to these MSW resonances.

Far from resonance, there is little suppression from the quantum Zeno effect.  This results from a coincidence in the functional form of the forward scattering potential and total scattering rate.  At temperatures much larger than the resonance temperature, the product $\Gamma_\alpha \ell_m$ approaches a constant much less than 1, while at temperatures much lower than the resonance temperature, $\Gamma_\alpha \ell_m$ rapidly falls as the temperature of the expanding universe decreases.

There is another possibility for maximizing the sterile neutrino production rate.  This occurs when the potential lepton number has decreased and/or the temperature has fallen to a point where the resonance condition, Eq.\ (\ref{eq:resonancecondition}), is no longer satisfied and there are no MSW resonances.  In this scenario, $\sin^2 2 \theta_M$ reaches a local maximum but is less than 1, which means that the neutrino oscillation length tends to be much smaller than in the resonant case.  Additionally, this occurs at later times, and thus lower temperatures, during the expansion of the universe, which means the product $\Gamma_\alpha \ell_m$ is significantly smaller than in the resonant case.  As a result, the overall sterile neutrino production rate can be higher than in the resonant case and would have a broader peak than in the resonant case.  With the right conditions, this nonresonant production of sterile neutrinos can become significant.

Throughout this section we have discussed the transformation properties of $\nu_\alpha \leftrightarrow \nu_s$ using the quantum kinetic equations and the quantum Zeno ansatz.  However, to consider the general problem of neutrino transformation in the early universe, we must also include its $CP$-counterpart, $\bar\nu_\alpha \leftrightarrow \bar\nu_s$.  The quantum kinetic equations and quantum Zeno ansatz are similar for the $\bar\nu_\alpha$, $\bar\nu_s$ system with a few alterations.  The forward scattering potential $V_\alpha$ [Eq.\ (\ref{eq:fwdscatteringpot})] is adjusted by replacing $\mathcal{L}_\alpha$ with $- \mathcal{L}_\alpha$, and the total scattering rate $\Gamma_\alpha$ [Eq.\ (\ref{eq:totscatteringrate})] is replaced by one appropriate for antineutrinos, $\bar\Gamma_\alpha$.  However in the early universe, to lowest order, these two scattering rates are equal, $\Gamma_\alpha \approx \bar\Gamma_\alpha$.

In this paper we will consider large lepton numbers which helps to avoid the necessity of considering both neutrinos and antineutrinos.  For a large positive lepton number, all of the action occurs in the neutrino sector.  The effective matter mixing angle for antineutrinos is significantly less than their vacuum mixing angles, while neutrinos may experience the MSW-like resonances discussed above.  For large negative lepton numbers, the same is true, but for the $CP$-counterpart.

\section{Calculations}
\label{sec:calculations}

Calculating the evolution of the active and sterile neutrino distribution functions remains a daunting task.  In addition to the quantum Zeno ansatz, Eq.\ (\ref{eq:quantumzenoansatz}), we must elucidate the lepton number evolution.  We assume that the active neutrinos are in thermal equilibrium in the early universe and that active-sterile neutrino conversion is the only flavor changing interaction.  The former assumption of thermal equilibrium is a good one before weak decoupling where the high temperature of the universe ($T \gtrsim 3 \,{\rm MeV}$) ensures that the weak interaction time scale is much shorter than a Hubble time, the time scale for the expansion of the universe.  The latter assumption amounts to assuming that there is no exotic physics other than active-sterile neutrino oscillations.  For the epoch before weak decoupling, the time evolution of the lepton number $L_\alpha$ is
\begin{equation}
\frac{d}{d t} L_\alpha (t) = - \frac{1}{4 \zeta(3)} \int_0^\infty d \epsilon \frac{\epsilon^2}{e^\epsilon + 1} \frac{\partial}{\partial t} f_s ( \epsilon, t ) .
\label{eq:leptonevolution}
\end{equation}

Together, the quantum Zeno ansatz [Eq.\ (\ref{eq:quantumzenoansatz})] and lepton number evolution [Eq.\ (\ref{eq:leptonevolution})] form a formidable system of integro-partial differential equations.  To simplify the situation, we assume that the effective statistical weight, $g$, is constant.  Strictly speaking,  this is not the case in the early universe because species fall out of thermal equilibrium as time progresses and the temperature decreases.  However, a well accepted detailed history $g(t)$ does not currently exist.  As a result, for illustrative purposes, we will assume that $g$ is constant.

By making this assumption we benefit in a number of ways.  The temperature evolution of the universe, $T(t)$, is greatly simplified.  As different species fall out of thermal equilibrium, particle-antiparticle pair annihilation reheats the universe, resulting in a complicated temperature evolution.  The temperature evolution, as given by the Friedman equation, the conservation of comoving entropy, and assuming radiation domination, is 
\begin{equation}
\frac{1}{T} \frac{d T}{d t} = - \left( \frac{4 \pi^3}{45} \right)^{1/2} g^{1/2} \frac{T^2}{m_{\rm PL}} .
\label{eq:temperatureevolution}
\end{equation}
Since $T(t)$ is a monotonically decreasing function of time, we can invert it to find $t(T)$.  We are then able to perform our calculations as a function of temperature.

In addition, a constant $g$ means that, by Eq.\ (\ref{eq:varepsilon}), $d \epsilon / d t = 0$.  As a result, the quantum Zeno ansatz becomes an ordinary differential equation for a family of sterile neutrino distribution functions, parameterized by $\epsilon$.  The only coupling between the different values of $\epsilon$ that remains is found in the lepton number evolution, Eq.\ (\ref{eq:leptonevolution}).

We choose a value of $g$ that is representative of the early universe before the QCD transition, \mbox{$170 \,{\rm MeV} \lesssim T \lesssim 1 \,{\rm GeV}$}.  In this epoch, the plasma of the early universe consists of photons ($g_b = 2$), gluons ($g_b = 16$), $e^\pm$ and $\mu^\pm$ pairs ($g_f = 8$), neutrinos and antineutrinos of all three active flavors ($g_f = 6$), and up, down, and strange quarks and antiquarks ($g_f = 36$), so $g \approx 61.75$, where we neglect corrections of order the lepton numbers and baryon to photon ratio.  At higher temperatures the other quarks come into the picture, but at this point the time scale for expansion in the early universe is so fast that there is little net effect from the difference in $g$ values.

We use the forward scattering potential given in Eq.\ (\ref{eq:fwdscatteringpot}), which is accurate at times after the QCD transition and for lepton numbers much larger than the baryon to photon ratio.  However, we are interested in times before the QCD transition where gluons and free quarks thermally populate the early universe.  The proper form of the forward scattering potential remains an open area of research \cite{bh07a, bh07b}.  The results presented in this paper will remain relevant unless the total forward scattering potential arising from the quarks is many orders of magnitude larger than the forward scattering on color singlet baryons and mesons with the same net baryon number.

On the other hand, the total scattering rate is greatly influenced by the additional species thermalized before the QCD transition.  However, since neutrino-quark interactions are not well studied, the total scattering rate in the QCD epoch remains a question mark.  We use the values given in Sec.\ \ref{sec:qke}, although our calculations deal with a different epoch in the early universe.  In addition, we calculate the effects of an increased scattering rate, by assuming that this rate is proportional to the number of relativistic particles with weak charge populated in the early universe.

We solve the system of integro-differential equations posed by the quantum Zeno ansatz for different values of $\epsilon$, and the lepton evolution.  Using Eq.\ (\ref{eq:temperatureevolution}), we change variables to follow the sterile neutrino distribution functions and lepton number as a function of the declining temperature of the expanding universe.  The initial conditions are chosen so that there are initially no sterile neutrinos, $f_s = 0$, and the initial lepton number is a free parameter, but we assume that there is an equivalent lepton number in each of the active neutrino species.  The initial (final) temperature for the calculations is chosen to be high (low) enough that the choice of this has no effect on the final outcome (sterile neutrino spectrum and final lepton number).  

\begin{figure*}
\includegraphics[width = 4in, angle = 270]{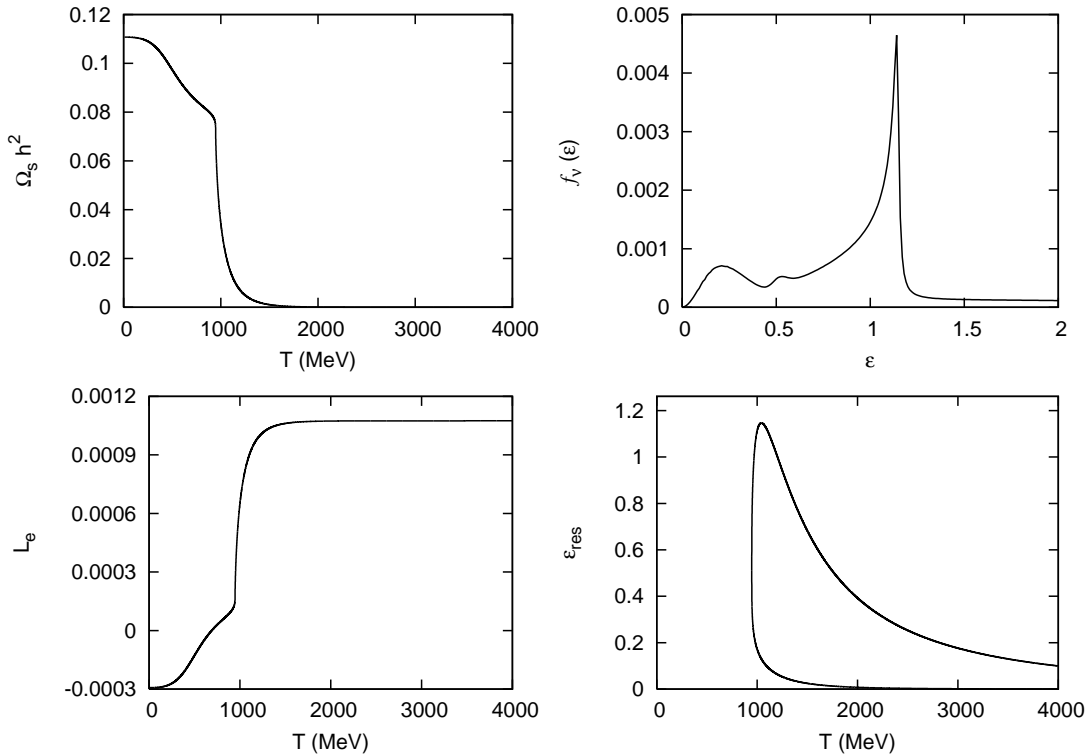}
\caption{\label{fig:ex} Results for our calculation using the quantum Zeno ansatz.  In this calculation, we take the sterile neutrino rest mass $m_s = 64 \,{\rm keV}$, vacuum mixing angle $\sin^2 2 \theta = 10^{-10}$, and initial lepton number $L_{e0} = 1.1 \times 10^{-3}$.  The upper left panel shows the cumulative sterile neutrino production history in terms of its closure fraction in the current epoch, $\Omega_s h^2$.  Below that, in the lower left corner, is the lepton number history.  The upper right panel displays the final $\nu_s$ spectrum as a function of scaled momentum $\epsilon$.  In the lower right panel is a plot that tracks the resonant values of $\epsilon$ as our calculation evolves in temperature.}
\end{figure*}

For illustrative purposes, we choose to work with electron neutrinos as the sole active species which mixes with sterile neutrinos ($\alpha = e$).  We do not include mixing with $\nu_\mu$ or $\nu_\tau$; this includes active-active mixing with $\nu_e$ and among themselves, and active-sterile mixing with $\nu_s$.    This is unrealistic, since the neutrino mixing matrix could be quite complicated.  However, our simplistic model will at least serve to illustrate trends in sterile neutrino production and lepton number depletion \cite{afp, abfw}.

In addition to the numerical procedure outlined above, we also solve the quantum kinetic equations, Eqs.\ (\ref{eq:qkePvec}$-$\ref{eq:qkeP0}), for a given value of $\epsilon$.  Using the lepton evolution calculated with the quantum Zeno ansatz, we compare the evolution of the sterile neutrino distribution function between the two methods.  The principal difficulty in solving the quantum kinetic equations is the need to resolve rapidly varying complex phases.  To ease these computational demands, we employ an eighth-order Runge-Kutta method \cite{rk8} which allows the use of a larger time step to solve the problem.

\section{Results}
\label{sec:results}

The particulars of phenomena revealed by the calculations described in the last section depend on the assumed values of sterile neutrino mass $m_s$, vacuum mixing angle $\theta$, and initial lepton number, $L_{e0}$, in each active neutrino species.  (As discussed above, we take $L_{e0} = L_{\mu 0} = L_{\tau 0}$.)  Some ranges of the sterile neutrino parameters can be ruled out by the x-ray observations \cite{aft, afp, dh02, ak06, bnos06, bnos06b, bnost06, pcss07, ybw07, wbyw06} or by observations of smaller-scale large scale structure ({\it e.g.}, the \mbox{Lyman-$\alpha$} forest) coupled with structure formation calculations \cite{pcss07, vlhmr05, a06, smmt06}.  Bounds on the lepton numbers  from big bang nucleosynthesis considerations \cite{kssw01, abb02, wong02, dhpprs02} are much larger than those employed in our calculations.  It is important to note that these lepton number limits constrain the final lepton numbers, not the initial lepton numbers.  However, other parameter values remain unconstrained by these considerations and, therefore, are still ``in play'' for providing a relic sterile neutrino density which could comprise some or all of the dark matter.  The observed dark matter closure fraction, $\Omega_c$, is measured to be \mbox{$\Omega_c h^2 = 0.1105 \pm 0.0039$} where $h \equiv H_0 / 100 \,{\rm km} \,{\rm s}^{-1} \,{\rm Mpc}^{-1}$, with $H_0$ the Hubble parameter \cite{wmap3}.

A set of parameters that produces the correct relic mass density to be the dark matter, but is nevertheless ruled out by x-ray observations provides particularly instructive cases.  These parameters give lepton number depletion and sterile neutrino production histories which show a variety of possible behaviors associated with resonant and nonresonant evolution.  Scenarios associated with these parameters give distinctive features in the final relic sterile neutrino energy spectrum which are directly attributable to the way the MSW resonance sweep couples with lepton number depletion.

Results for an example set of these parameters are shown in Fig.\ \ref{fig:ex}.  For this case we have taken $m_s = 64 \,{\rm keV}$, $\sin^2 2 \theta = 10^{-10}$, and $L_{e0} = 1.1 \times 10^{-3}$.  Using the quantum Zeno ansatz, our calculations yield a sterile neutrino relic density which is consistent with the observed dark matter closure fraction.  

The panel in the lower right of \mbox{Fig.\ \ref{fig:ex}} shows the sweep history of the two MSW resonances ({\it i.e.}, as in Eq.\ \ref{eq:mswresonancesoln}).  The resonant scaled momentum ($\epsilon_{\rm res}$) of each resonance has a characteristic behavior.  As the universe expands, the temperature $T$ drops, and lepton number is depleted, these resonances first diverge and then converge in scaled momentum.  At a temperature $\sim 950 \,{\rm MeV}$, the temperature has fallen and the lepton number has been depleted to a point where the resonance condition, Eq.\ (\ref{eq:resonancecondition}), cannot be satisfied and so subsequently there can be no MSW resonances.

The lepton number history and the cumulative sterile neutrino production history (as measured by its closure fraction in the current epoch $\Omega_s h^2$) are shown in the lower and upper left-hand panels of Fig.\ \ref{fig:ex}, respectively.  The temperature where MSW resonances cease to exist corresponds to an abrupt shift in the sterile neutrino production mechanism, from resonant to nonresonant production.  Roughly 70 percent of the total relic sterile neutrino density is produced resonantly.  At temperatures slightly above those characteristic of the resonance cessation event, there is an epoch of precipitous depletion of lepton number and concomitant production of sterile neutrinos.  

The sharp decline in lepton number and the subsequent loss of resonance are a consequence of the coupling between the sweep of the MSW resonances and lepton number evolution.  As the MSW resonances sweep through the $\nu_e$ distribution, and $\nu_e$'s are converted to $\nu_s$'s, lepton number is lost.  In turn, the decreasing lepton number accelerates the resonance sweep rate.  This positive feedback loop results in the precipitous depletion of lepton number.  This process ends once MSW resonances cease to exist, as seen in our calculations.  

This behavior, coupled with the dominance of resonant $\nu_s$ production, results in a characteristic set of peaks in the relic sterile neutrino energy spectrum 
\begin{equation}
f_\nu (\epsilon) \equiv f_s (\epsilon, T_f) \frac{\epsilon^2}{e^\epsilon + 1},
\end{equation}
where $T_f$ is the final temperature in the calculations.  These features are seen in the upper right panel of Fig.\ \ref{fig:ex}.  In this figure, there are two distinct peaks seen at $\epsilon \approx 0.2$ and $\epsilon \approx 1.15$, and a smaller peak at $\epsilon \approx 0.5$.  The three peaks are the result of 
different physical processes.  Understanding these processes provides physical insight into the mechanism of sterile neutrino production.

Since the magnitude of the lepton number is small throughout the calculation, $\vert L_e \vert \ll 1$, we can make the approximation that $f_e (\epsilon, T) \approx 1$, neglecting corrections of order the lepton number.  As a result, Eq.\ (\ref{eq:quantumzenoansatz}) can be solved analytically if we use the lepton number evolution $L_e (T)$ from our calculations.  This will provide an analytical backbone upon which to base our analysis.  

Using this approximation, the solution to the quantum Zeno ansatz yields a relic sterile neutrino energy spectrum
\begin{equation}
f_\nu (\epsilon) \approx \frac{\epsilon^2}{e^\epsilon + 1} \left( 1 - \exp \left\{ - \int_{T_0}^{T_f} \gamma (\epsilon, \tau) \,d \tau \right\} \right),
\end{equation}
where 
\begin{equation}
\gamma (\epsilon, T) \equiv \frac{1}{4\dot{T} (T)} \frac{\Gamma_e (\epsilon, T) \sin^2 2 \theta_M (\epsilon, T)}{1 + \frac{1}{4} \Gamma_e^2 (\epsilon, T) \ell_m^2 (\epsilon, T)}
\end{equation}
is the $\nu_e \leftrightarrow \nu_s$ conversion rate per unit temperature interval.  In theory, the initial temperature $T_0 \rightarrow \infty$, but in practice is chosen as described in Sec.\ \ref{sec:calculations}.

For a fixed value of $\epsilon$, $\gamma(\epsilon, T)$ is a sharply peaked function of temperature $T$ whenever MSW resonances exist.  This is the result of this function's proportionality to $\sin^2 2 \theta_M$, which has sharp peaks when $\sin^2 2 \theta_M = 1$ at the MSW resonances discussed above.  The locations in temperature of these resonances are implicitly given by the solutions of Eq.\ (\ref{eq:mswresonanceeqn}) for a fixed value of $\epsilon$.  

However, the widths of the peaks in $\gamma(\epsilon, T)$ do not correspond to the widths of the associated MSW resonances.  Both $\sin^2 2 \theta_M (\epsilon, T)$ and $\ell_m^2 (\epsilon, T)$ are sharply peaked functions.  The resonance width of both functions is  $\delta T \approx \mathcal{H} \tan 2 \theta$ (see, {\it e.g.} Ref.\ \cite{abfw}), where $\mathcal{H}$ is the density scale height at resonance and is defined as
\begin{eqnarray}
\mathcal{H} &\equiv &\left\vert \frac{1}{V_e} \frac{d V_e}{d T} \right\vert^{-1}_{\rm res} \\
 & = &\left\vert \frac{5}{T} - \frac{8 \sqrt{2} \zeta (3)}{\pi^2 \delta m^2 \cos 2 \theta} G_F \epsilon \mathcal{L} T^3 \left( 1 - \frac{d \log \mathcal{L}}{d \log T} \right) \right\vert^{-1}_{\rm{ res}}. \nonumber
\end{eqnarray}
In the neighborhood of each resonance, $\Gamma_e^2 \ell_m^2 / 4 \gg 1$, so near resonance 
\begin{equation}
\gamma (\epsilon, T) \approx \frac{1}{\dot{T} (T) \Gamma_e (\epsilon, T)} \frac{\sin^2 2 \theta_M (\epsilon, T)}{\ell_m^2 (\epsilon, T)} .
\end{equation}
The sharp peaks in both $\sin^2 2 \theta_M (\epsilon, T)$ and $\ell_m^2 ( \epsilon, T )$, in a sense, cancel each other out.  This is because the ratio of these functions,
\begin{equation}
\frac{\sin^2 2 \theta_M (\epsilon, T)}{\ell_m^2 (\epsilon, T)} = \left( \frac{\delta m^2 \sin 2 \theta}{2 \epsilon T} \right)^2 ,
\end{equation}
is no longer strongly peaked.

A measure of the width of the resonances in $\gamma(\epsilon, T)$ can be obtained by setting $\Gamma_e^2 \ell_m^2 / 4 = 1$. Far from resonance, $\Gamma_e^2 \ell_m^2 / 4 \lesssim 1$, so that $\gamma (\epsilon, T) \propto \epsilon \,T^2 \sin^2 2 \theta_M$, which means $\gamma(\epsilon, T)$ has the same resonant shape as $\sin^2 2 \theta_M$.  When $\Gamma_e^2 \ell_m^2 / 4 \gtrsim 1$ closer to the resonance, $\gamma (\epsilon, T) \propto \epsilon^{-3} T^{-10}$, which means the peaked behavior is truncated and the location of the maximum of the neutrino conversion rate is displaced from the location of the MSW resonance to a slightly lower temperature.  The change in the forward scattering potential required to satisfy this condition is $\delta V_e \approx \Gamma_e / 2$.  The width in temperature space corresponding to this potential width is
\begin{eqnarray}
\delta T (\epsilon, T_{\rm res}) & = & \left\vert \frac{\delta V_e}{V_e} \right\vert_{\rm res} \left\vert \frac{1}{V_e} \frac{d V_e}{d T} \right\vert_{\rm res}^{-1} \nonumber \\
 & \approx & \frac{\epsilon \,T_{\rm res}}{\delta m^2 \cos 2 \theta} \Gamma_e (\epsilon, T_{\rm res}) \mathcal{H}.
\end{eqnarray}

If we restrict ourselves to values of $\epsilon$ for which MSW resonances are possible ($\epsilon < 1.15$ for the calculations done with the parameters in Fig.\ \ref{fig:ex}), then the sharp peaks in $\gamma (\epsilon, T)$ allow us to make the approximation:
\begin{equation}
\int_{T_0}^{T_f} \gamma (\epsilon, \tau) \,d \tau \approx \sum_i g_i \gamma ( \epsilon, T_{{\rm res,}i} ) \delta T (\epsilon, T_{{\rm res,}i}),
\label{eq:gammadT}
\end{equation}
where $g_i$ is a numerical coefficient of order unity, $T_{{\rm res,}i}$ is the temperature of the MSW resonance for scaled momentum $\epsilon$, and the sum is performed over every resonance.  This approximation breaks down when two resonances begin to overlap, a situation we will discuss below.

The product
\begin{equation}
\gamma (\epsilon, T_{\rm res}) \delta T \approx \frac{1}{2} \frac{\delta m^2}{2 \epsilon T_{\rm res}} \frac{\sin^2 2 \theta}{\cos 2 \theta} \left( \dot{T}^{-1} \mathcal{H} \right),
\end{equation}
is equal to one-half of the the dimensionless adiabaticity parameter defined in Eq.\ (38) of Ref.\ \cite{abfw}, but with the flavor off-diagonal potential set to zero, $B_{e\tau} = 0$.  Note also that the density scale height in this work, $\mathcal{H}$, is defined as the width in temperature of the MSW resonance, but the density scale height in Ref.\ \cite{abfw}, $\mathcal{H} = \dot{T}^{-1} \mathcal{H}$, is defined as the physical width of the MSW resonance. 

If the $g_i$ in Eq.\ (\ref{eq:gammadT}) are equal to $\pi$, then this result is equivalent to the Landau-Zener approximation.  However, these numerical coefficients depend on the shape of the resonances, in particular on the shape of the forward scattering potential $V_e$.  On the other hand, the coherent $\nu_e \rightarrow \nu_s$ probability is only equal to the Landau-Zener approximation when the forward scattering potential is linear.  It is interesting to see the relationship between the coherent $\nu_e \rightarrow \nu_s$ conversion through the coherent MSW process and scattering-induced decoherent $\nu_s$ production.  Reference \cite{sf99} used this result without this analysis.

The low-$\epsilon$ peak is the product of two opposing factors:  a $\nu_e$ spectrum that increases and a $\nu_e \rightarrow \nu_s$ conversion probability that decreases as $\epsilon$ increases.  At small values of $\epsilon$, the number density of $\nu_e$ with scaled momentum $\epsilon$ that are available to be converted into $\nu_s$ is proportional to $\epsilon^2$.  

The lower right panel in Fig.\ \ref{fig:ex} shows that at these $\epsilon$ values there are two resonances, one at a high temperature ($\gtrsim 2000 \,{\rm MeV}$) and another at a lower temperature near the resonance cessation event ($\sim 1000 \,{\rm MeV}$).  At these temperatures $\delta T \propto \epsilon \,T^3 \vert d \log \mathcal{L} / d \log T - 1 \vert^{-1}$, so that the product $\gamma (\epsilon, T_{\rm res}) \delta T$ in Eq.\ (\ref{eq:gammadT}) is proportional to $\epsilon^{-2} T_{\rm res}^{-7} \vert d \log \mathcal{L} / d \log T - 1 \vert^{-1}$.   Consequently, the contribution from the low-temperature resonance is dominant over the contribution from the high-temperature resonance.  At small values of $\epsilon$ ($\epsilon \lesssim 0.4$), the relic sterile neutrino distribution is
\begin{equation}
f_\nu (\epsilon) \approx \frac{\epsilon^2}{e^\epsilon + 1} \left( 1 - \exp \left\{ - \kappa \epsilon^{-2} T_{\rm res}^{-7} \left\vert \frac{d \log \mathcal{L}}{d \log T} - 1 \right\vert^{-1}_{\rm res} \right\} \right),
\end{equation}
where $\kappa$ is a constant that depends on sterile neutrino parameters and physical constants, and $T_{\rm res} = T_{\rm res} (\epsilon)$ is the low-temperature solution to Eq.\ (\ref{eq:mswresonanceeqn}).  This distribution can be seen graphically in the lower right panel of Fig.\ \ref{fig:ex}.

Examination of Fig.\ \ref{fig:ex} reveals that $T_{\rm res} (\epsilon)$ is a slowly decreasing function of  $\epsilon$ ({\it e.g.}, a factor of 4 increase in $\epsilon$ leads to a 15 percent decrease in $T_{\rm res}$).  As a result, $\epsilon^{-2} T_{\rm res}^{-7}$ remains a decreasing function of $\epsilon$.  In addition, $d \log \mathcal{L} / d \log T$ is maximal immediately preceding the resonance cessation event and decreases with increasing temperature.  It follows that $\vert d \log \mathcal{L} / d \log T - 1 \vert^{-1}$ is also a decreasing function of $\epsilon$.  Thus, the cumulative $\nu_e \rightarrow \nu_s$ conversion probability, $1 - e^{- \int \gamma \,d \tau}$, is a decreasing function of $\epsilon$.  The interplay between this and the increasing spectrum of $\nu_e$ produces the peak seen at low values of $\epsilon$.

The high-$\epsilon$ peak corresponds to the maximum value of $\epsilon$ ($\epsilon_{\rm max}$) for which MSW resonances exist.  At this value of $\epsilon$, the MSW resonance equation, Eq.\ (\ref{eq:mswresonanceeqn}), has only one positive temperature solution (a double root).  In the case shown in Fig.\ \ref{fig:ex}, this occurs at $\epsilon_{\rm max} \approx 1.15$.  

For $\epsilon < \epsilon_{\rm max}$, the energy spectrum increases as $\epsilon$ increases, and as it approaches the maximum value, it increases at a greater rate until it hits the peak.  Here, the proportionality to $\epsilon$ loses the sway it had at lower values of $\epsilon$ because the fractional changes are an order of magnitude smaller.  In this regime, the high-temperature resonance moves to lower temperatures where it becomes significant in contributing to the $\nu_e \rightarrow \nu_s$ conversion probability.  Additionally, as the two resonances approach each other, the peaks tend to blend with each other, enhancing the value of $\int \gamma \,d \tau$, culminating with the merger of the two peaks.  The sharp increase in $f_\nu$ indicates the point at which this phenomenon becomes significant.

For values of $\epsilon$ slightly larger than $\epsilon_{\rm max}$, the energy spectrum falls off precipitously.  This is because neutrinos with $\epsilon > \epsilon_{\rm max}$ have no resonances, and thus sterile neutrino production is suppressed compared to the regime with resonant production.

The mid-$\epsilon$ peak corresponds to the last resonant value of $\epsilon$.  At the temperature where MSW resonances cease to exist, there is an abrupt transition from resonant to nonresonant sterile neutrino production.  During the final stages of resonant sterile neutrino production, there is a large conversion rate of active neutrinos to sterile neutrinos, producing a large lepton number loss rate, leading to a smaller resonance width.  Immediately after the resonance cessation event, lepton number is depleted at a far lower rate, as only nonresonant conversion remains.  This leads to the low-temperature side of the resonance being much broader than the high-temperature side.  As a result, there is an enhancement in the $\nu_e \rightarrow \nu_s$ conversion probability that is maximized at the last resonant $\epsilon$.  This effect is significantly smaller than the effects associated with the two peaks discussed previously.

\begin{figure*}
\includegraphics[width = 4in, angle = 270]{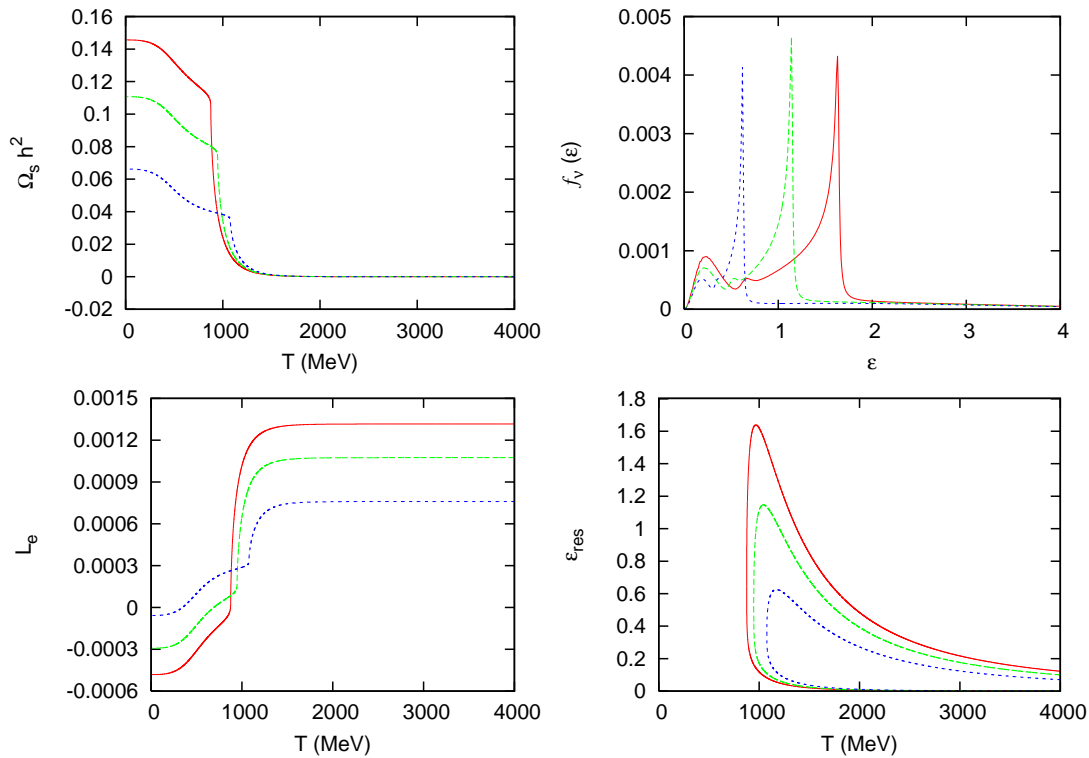}
\caption{\label{fig:lept} The four panels present the same quantities as in Fig.\ \ref{fig:ex}.  In this figure, we have taken $m_s = 64 \,{\rm keV}$, $\sin^2 2 \theta = 10^{-10}$, and three different values for the initial lepton number:  $L_{e0} = 0.76 \times 10^{-3}$ (short dashed curves), $1.1 \times 10^{-3}$ (long dashed curves), and $1.3 \times 10^{-3}$ (solid curves).}
\end{figure*}

\begin{figure*}
\includegraphics[width = 4in, angle = 270]{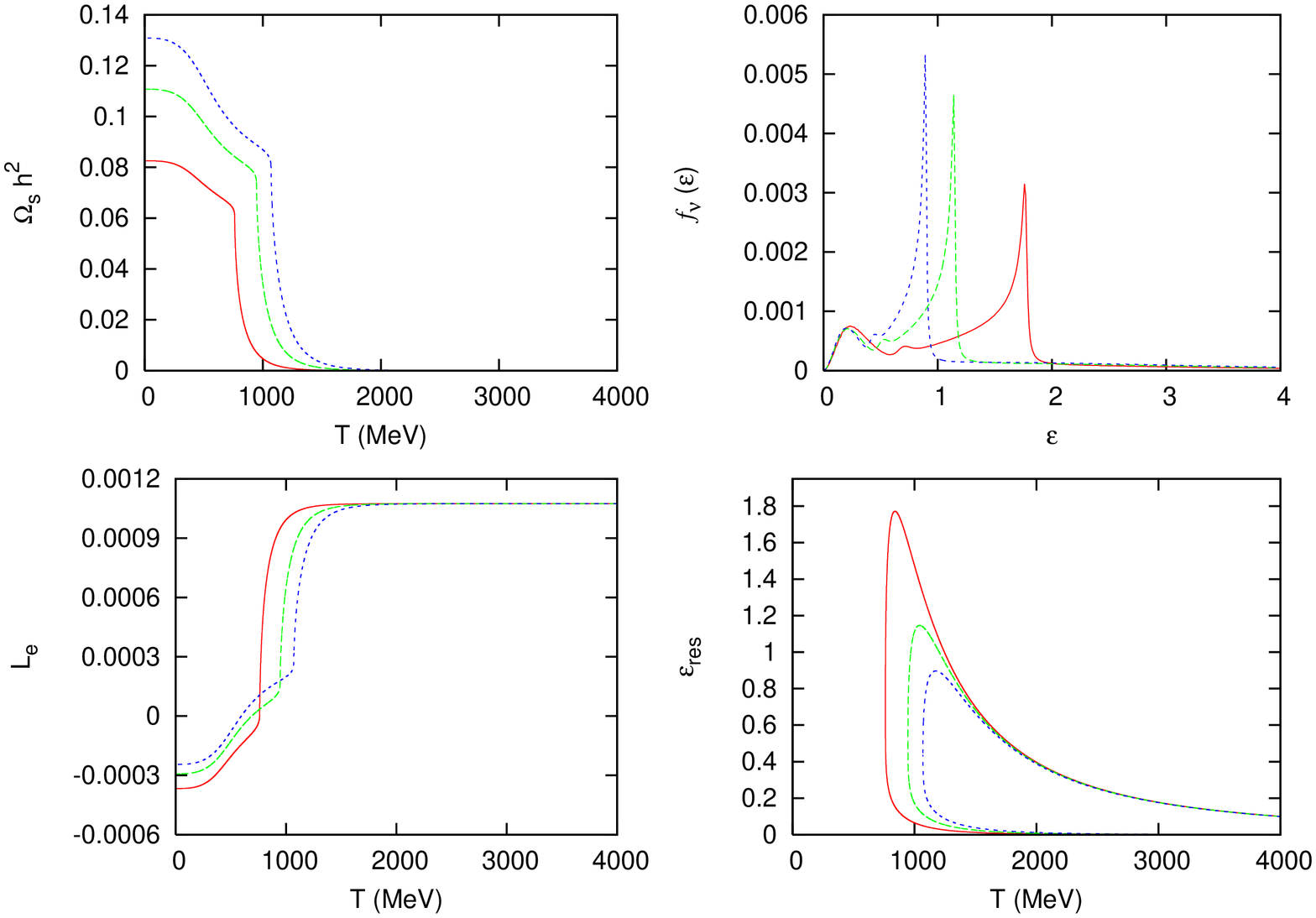}
\caption{\label{fig:mass} The four panels present the same quantities as in Fig.\ \ref{fig:ex}.  In this figure, we have taken $L_{e0} = 1.1 \times 10^{-3}$, $\sin^2 2 \theta = 10^{-10}$, and three different values for the sterile neutrino rest mass:  $m_s = 45 \, {\rm keV}$ (solid curves), $64 \, {\rm keV}$ (long dashed curves), and $79 \,{\rm keV}$ (short dashed curves).}
\end{figure*}

\begin{figure*}
\includegraphics[width = 4in, angle = 270]{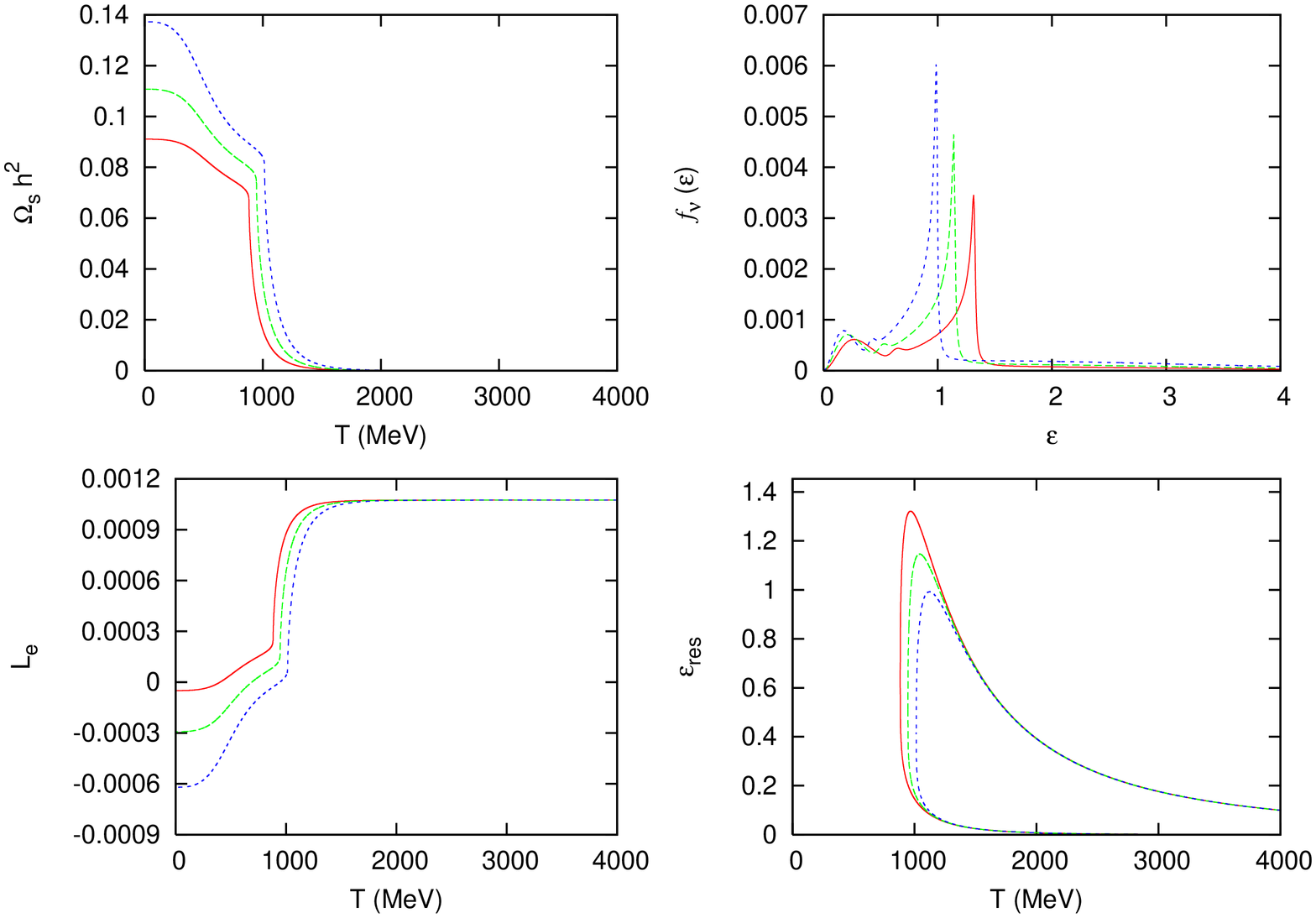}
\caption{\label{fig:sin2t} The four panels present the same quantities as in Fig.\ \ref{fig:ex}.  In this figure, we have taken $m_s = 45 \,{\rm keV}$, $L_{e0} = 1.1 \times 10^{-3}$, and three different values for the vacuum mixing angle:  $\sin^2 2 \theta = 0.5 \times 10^{-10}$ (solid curves), $1.0 \times 10^{-10}$ (long dashed curves), and $2.0 \times 10^{-10}$ (short dashed curves).}
\end{figure*}

Figures \ref{fig:lept} - \ref{fig:sin2t} show the effects of changing each of the three free parameters in our calculation.  Figure\ \ref{fig:lept} presents the effects of changing the initial lepton number in each active neutrino species, but leaving the sterile neutrino mass and vacuum mixing angle fixed.  Figure \ref{fig:mass} shows the effects of changing the sterile neutrino mass, and Figure\ \ref{fig:sin2t} shows the effects of changing the vacuum mixing angle $\theta$.  Each figure displays the same four panels as discussed for Fig. \ref{fig:ex}.  The results discussed above serve as a guide to understanding the trends in these figures.

In Fig.\ \ref{fig:lept}, we have taken $m_s = 64 \,{\rm keV}$ and $\sin^2 2 \theta = 10^{-10}$.  We have varied the initial lepton number and show three cases, corresponding to $L_{e0} = 0.76 \times 10^{-3}$, $1.1 \times 10^{-3}$, and $1.3 \times 10^{-3}$.  

The resonance cessation event occurs at a higher temperature for lower initial lepton numbers.  We can understand this result using the resonance condition, Eq.\ (\ref{eq:resonancecondition}).  Since we assume that initially the lepton number is equal in each active neutrino species and we only allow $\nu_e$-$\nu_s$ neutrino oscillations, the potential lepton number can be written as $\mathcal{L} (T) = 2 L_{e0} + 2 L_e (T)$.  Most of the initial $\nu_e$ lepton number is depleted by $\nu_e \rightarrow \nu_s$ conversion, so that at the cessation event $\mathcal{L} \approx 2 L_{e0}$.  As a result, lower initial lepton numbers mean an earlier (higher temperature) loss of resonance.

The maximum resonant $\epsilon$, $\epsilon_{\rm max}$, increases with increasing initial lepton number.  As seen in Eq.\ (\ref{eq:mswresonancesoln}), the upper resonance is proportional to $\mathcal{L} / T^2$.  This implies that the resonant scaled momenta can reach higher values for higher initial lepton numbers.  For $\epsilon \lesssim 2.2$ there are more $\nu_e$ available to convert into $\nu_s$ at a given value of $\epsilon$ as $\epsilon$ increases.  

We must perform the same analysis done above in estimating the cumulative $\nu_e \rightarrow \nu_s$ conversion probability.  Preserving the proportionality to the potential lepton number and sterile neutrino mass and mixing angle, we have
\begin{equation}
\gamma (\epsilon, T_{\rm res}) \delta T \propto \epsilon^{-2} T_{\rm res}^{-7} \left\vert \frac{d \log \mathcal{L}}{d \log T} - 1 \right\vert^{-1}_{\rm res} \mathcal{L}^{-1} m_s^4 \sin^2 2 \theta ,
\label{eq:gammadTprop}
\end{equation}
where the assumed sterile neutrino mass is $m_s \approx \sqrt{\delta m^2}$.  The relevant resonance temperatures are approximately equal to the temperature of the resonance cessation event, which increases with decreasing initial lepton number.  In our calculations, $T_{\rm res}^{-7}$ dominates over $\mathcal{L}^{-1}$, so in concert with the trend for $\epsilon_{\rm max}$,  increasing the initial lepton number leads to a larger (in magnitude) loss of $\nu_e$ lepton number,
\begin{equation}
\Delta L_e = - \frac{1}{4 \zeta (3)} \int_0^\infty f_\nu (\epsilon) \,d \epsilon .
\end{equation}
Consequently, the number density of sterile neutrinos produced, $n_{\nu_s} = - n_\gamma \Delta L_e$, increases with increasing initial lepton numbers.  Thus, the sterile neutrino closure fraction, $\Omega_s \propto m_s n_{\nu_s}$, also increases with increasing initial lepton number.

In Fig.\ \ref{fig:mass}, we have taken $L_{e0} = 1.1 \times 10^{-3}$, \mbox{$\sin^2 2 \theta = 10^{-10}$} and varied the sterile neutrino rest mass.  In this figure we show the results of our calculations for sterile neutrino masses $m_s = 45 \,{\rm keV}$, $64\,{\rm keV}$, and $79\,{\rm keV}$.

It is clear from this figure that the resonance cessation event occurs at higher temperatures for higher masses.  This trend can be understood by noticing that the right hand side of Eq.\ (\ref{eq:resonancecondition}) is proportional to $m_s$.  Therefore, with the same arguments as above, but now with a fixed lepton number, equality in Eq.\ (\ref{eq:resonancecondition}) (which marks the end of resonances) is reached at higher temperatures for higher assumed sterile neutrino masses.

In this calculation, at high temperatures \mbox{($T \gtrsim 2000 \,{\rm MeV}$)} the resonant values of $\epsilon$ do not depend on sterile neutrino mass.  Using the proportionality in Eq.\ (\ref{eq:gammadTprop}), lepton number depletion, and the concomitant sterile neutrino production, will be highest for the higher mass sterile neutrinos, at least early on.  As a result, the highest mass sterile neutrino will have the lowest value of $\epsilon_{\rm max}$ and the highest temperature at which the resonance cessation event occurs.  

The two competing factors in Eq.\ (\ref{eq:gammadTprop}), $T_{\rm res}^{-7}$ and $m_s^4$, nearly balance each other, but the higher value of $\epsilon_{\rm max}$ for lower sterile neutrino masses results in a larger (in magnitude) value of $\Delta L_e$.  However, it can be seen in the lower left panel of Fig.\ \ref{fig:mass} that the values of $\Delta L_e$ for the three sterile neutrino masses shown do not differ by much.  As a result, the sterile neutrino closure fraction, the relevant quantity in searching for a dark matter candidate, increases for increasing assumed sterile neutrino mass.

In Fig.\ \ref{fig:sin2t}, we have taken $m_s = 64 \,{\rm keV}$ and $L_{e0} = 1.1 \times 10^{-3}$ but have employed different values of the vacuum mixing angle $\theta$.  The figure shows three cases, for $\sin^2 2 \theta = 0.5 \times 10^{-10}$, $1.0 \times 10^{-10}$, and $2.0 \times 10^{-10}$.

For the vacuum mixing angles that we are concerned with here, the locations of the MSW resonances are insensitive to the value of $\theta$ (since $\cos 2 \theta \approx 1 - 0.5 \sin^2 2 \theta$).  As a result, differences in the evolution of $\epsilon_{\rm res}$ are linked to the differences in the lepton number evolution.  At high temperatures ($T \gtrsim 1500 \,{\rm MeV}$ for this calculation), the resonances sweep through the $\nu_e$ distribution at the same rate, but the proportionality to $\sin^2 2 \theta$ in Eq.\ (\ref{eq:gammadTprop}) leads to a larger sterile neutrino production for larger vacuum mixing angles.  As a result, the resonance condition, Eq.\ (\ref{eq:resonancecondition}), will be violated at the highest temperature for the highest mixing angle.  Thus, the largest mixing angle corresponds to the highest resonance cessation temperature and the smallest value of $\epsilon_{\rm max}$.

The combination of the proportionality of the cumulative $\nu_e \rightarrow \nu_s$ conversion rate to $\sin^2 2 \theta$, along with the weak dependence of the resonant temperatures and $\epsilon_{\rm max}$ on the vacuum mixing angle, result in resonant production of sterile neutrinos being highest for higher values of vacuum mixing angle.  However, this is just the tip of the iceberg.  Nonresonant scattering-induced decoherent production of sterile neutrinos is also proportional to $\sin^2 2 \theta$.  As a result, higher vacuum mixing angles generally result in a higher sterile neutrino closure fraction.

\begin{figure*}
\includegraphics[width = 4in, angle = 270]{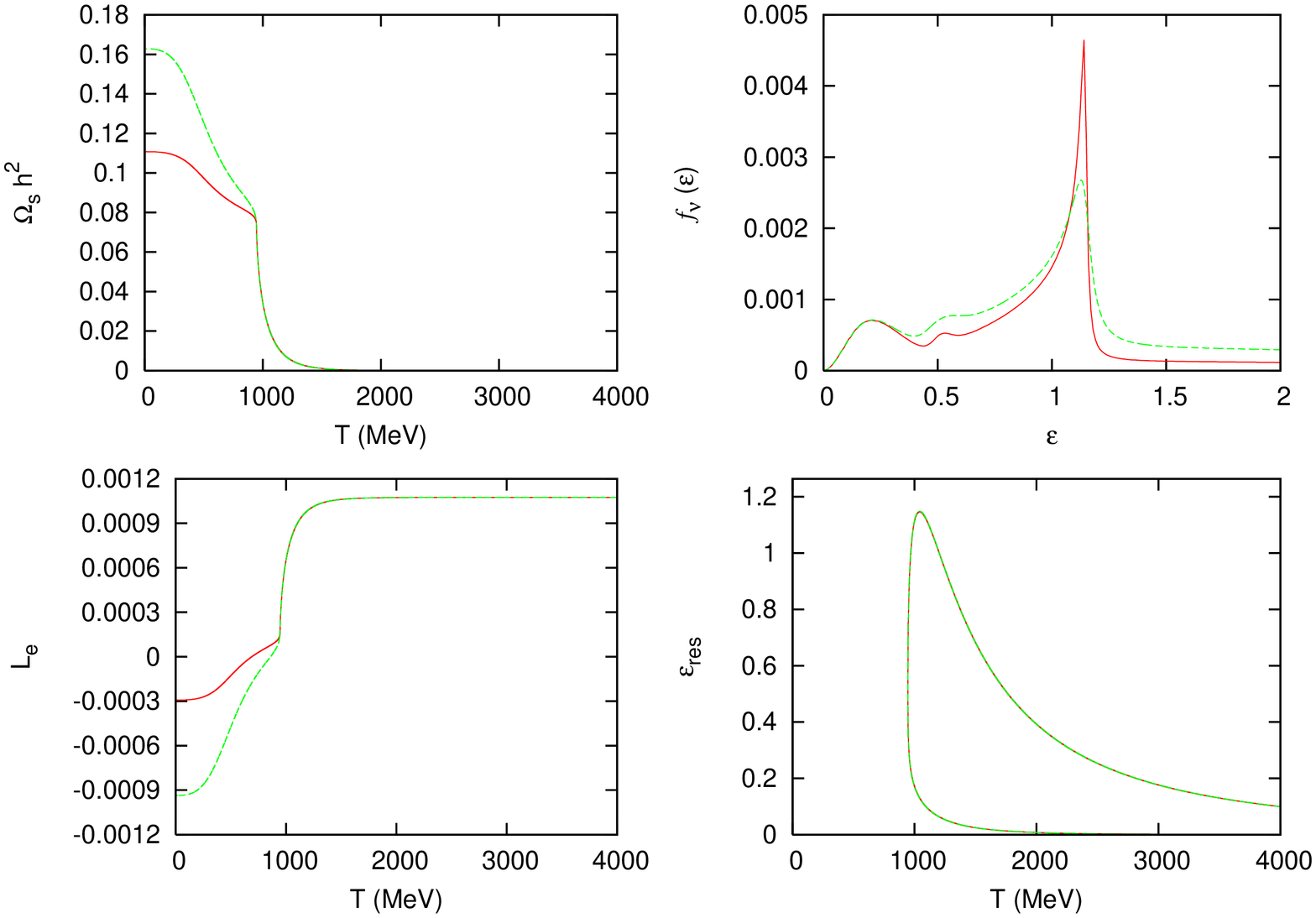}
\caption{\label{fig:hiG} The four panels present the same quantities as in Fig.\ \ref{fig:ex}.  In this figure, we have taken $m_s = 64 \,{\rm keV}$, $L_{e0} = 1.1 \times 10^{-3}$, and $\sin^2 2 \theta = 10^{-10}$.  We present two calculations, one with a low total scattering rate $\Gamma_e$ as discussed in section \ref{sec:qke} (solid curves), and one with a higher total scattering rate $\Gamma_e \rightarrow 3.5 \Gamma_e$ (dashed curves).  For $T \gtrsim 950 \,{\rm MeV}$, the cumulative sterile neutrino production history, lepton number history, and resonant values of $\epsilon$ are nearly identical in the two calculations.}
\end{figure*}

In Fig.\ \ref{fig:hiG}, we adopt the same parameters as in Fig.\ \ref{fig:ex} ($m_s = 64 \,{\rm keV}$, $\sin^2 2 \theta = 10^{-10}$, and $L_{e0} = 1.1 \times 10^{-3}$), but we modify the calculation by increasing the total scattering rate, $\Gamma_e$.  In the calculations for Figs.\ \ref{fig:ex} - \ref{fig:sin2t}, we have used a value of $y_e = 1.27$ in the total scattering rate, Eq.\ (\ref{eq:totscatteringrate}).  However, as discussed in Sec.\ \ref{sec:qke}, this is a value that is relevant for temperatures much lower than those in our calculations.  

We modify our calculations by assuming that the total scattering rate is proportional to the number of weak degrees of freedom that are in thermal equilibrium in the early universe.  The value $y_e \approx 1.27$ stems from $\nu_e$ scattering on $e^\pm$ pairs and neutrinos and antineutrinos of all three active flavors.  However, for the temperatures that we are concerned with here, there are also $\mu^\pm$ pairs and up, down, and strange quarks and antiquarks thermally populated in the early universe.  By including all three quark colors and the fact that the weak current is left-handed, we modify the total scattering rate by a factor of 3.5.  In other words, in testing the sensitivity to the scattering rate, we substitute $\Gamma_e (\epsilon, T) \rightarrow 3.5 \Gamma_e (\epsilon, T)$.

This calculation shows that there is little effect in the resonant conversion regime.  This result can be understood by noticing that $\gamma(\epsilon, T_{\rm max}) \propto \Gamma_e^{-1}$ and $\delta T \propto \Gamma_e$.  Therefore, the resonant production of sterile neutrinos is only weakly dependent on the total scattering rate.  In Fig.\ \ref{fig:hiG} this is readily apparent, because the evolution of the resonant scaled momentum and the lepton number (and consequently the sterile neutrino production history) are nearly identical for $T \gtrsim 950 \,{\rm MeV}$. 

The most dramatic effect of an increased scattering rate can be seen in the sterile neutrino production history after the resonance cessation event.  Roughly 55percent of the final sterile neutrino relic density is produced nonresonantly (compared with 30 percent in the lower $\Gamma_e$ case).  This is because away from resonance the quantum Zeno effect is less efficient at suppressing sterile neutrino production, allowing the $\nu_e \rightarrow \nu_s$ rate to increase with $\Gamma_e$.

Another effect of the increased scattering rate is seen in the sterile neutrino energy distribution function.  The most significant difference is that the high-$\epsilon$ peak is not as sharply peaked in this case.  This is due to the combination of two processes.  Nonresonant sterile neutrino production for $\epsilon > \epsilon_{\rm max}$ is enhanced with a larger scattering rate, producing a less steep drop off.  The merging of the two resonances that created the sharp peak in the low $\Gamma_e$ case is less effective for a larger $\Gamma_e$.  As discussed above, the peaks are broader with a smaller amplitude.  Consequently, when these peaks come together, the result is a smaller cumulative sterile neutrino production.

\begin{figure}
\includegraphics[width = 2in, angle = 270]{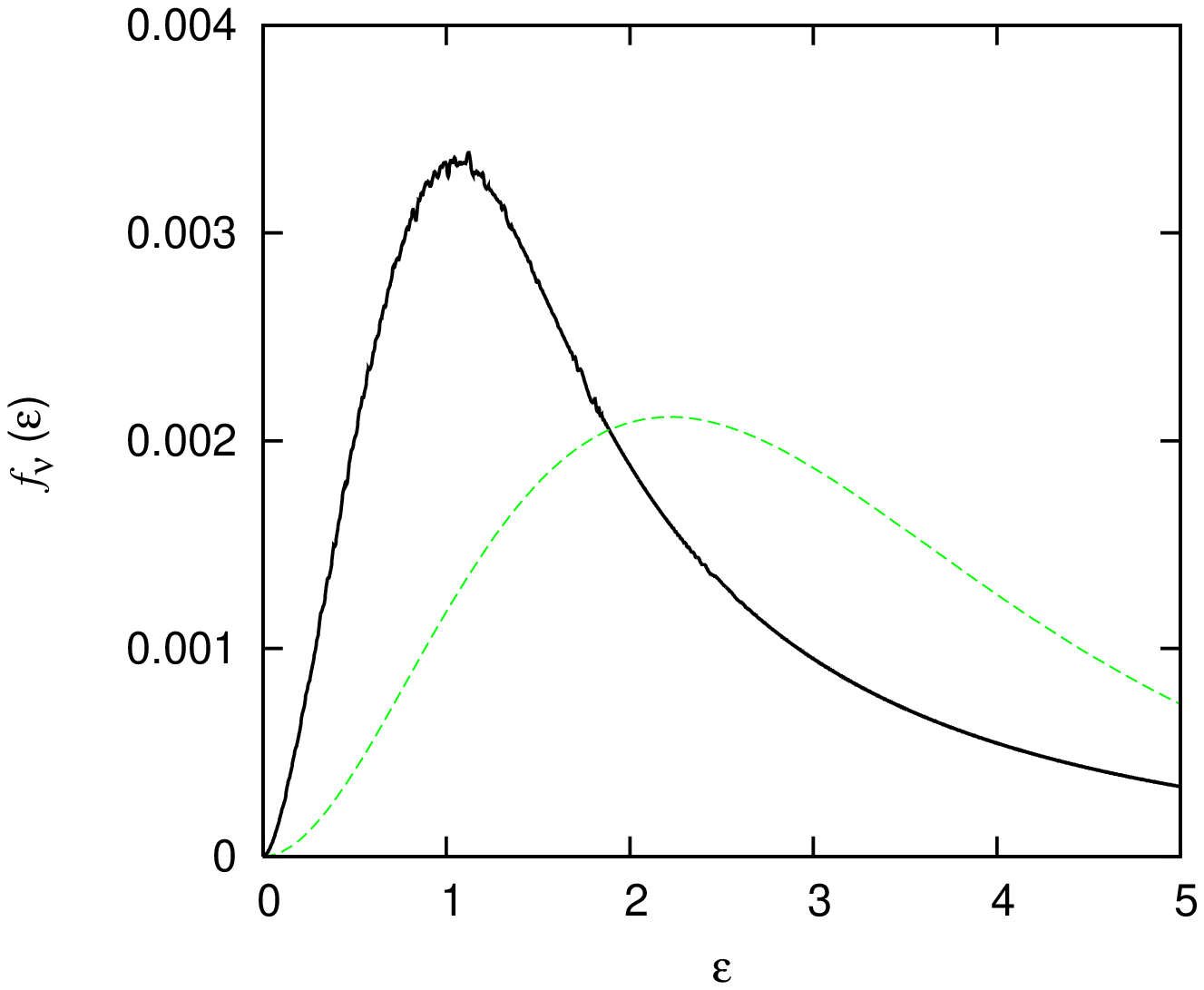}
\caption{\label{fig:inplay} The sterile neutrino energy spectrum as a function of scaled momentum $\epsilon$ for an ``in play'' sterile neutrino dark matter candidate with $m_s = 12 \,{\rm keV}$, $\sin^2 2 \theta = 3.2 \times 10^{-11}$, and $L_{e0} = 4.8 \times 10^{-3}$ (solid curve).  Also shown is a Fermi-Dirac spectrum normalized so that both spectra would produce the same number density of sterile neutrinos (dashed curve).}
\end{figure}

To this point we have only discussed sterile neutrino dark matter candidates that are ruled out by the current X-ray observations.  In Fig.\ \ref{fig:inplay} we present the energy spectrum for a sterile neutrino dark matter candidate that is still ``in play'' as far as the x-ray observations are concerned.  To evade the x-ray limits, we must reduce the sterile neutrino mass and/or vacuum mixing angle.  From the studies we conducted above, we know that the relic sterile neutrino closure fraction is more strongly dependent on mass than on mixing angle.  Therefore, we must reduce both mass and mixing angle to get under the x-ray limits.  However, to get the correct relic density we also increase the initial lepton number and the scattering rate as discussed above for Fig.\ \ref{fig:hiG}. 

For this calculation we use a sterile neutrino rest mass $m_s = 12 \,{\rm keV}$ with  vacuum mixing angle $\sin^2 2 \theta = 3.2 \times 10^{-11}$ and initial lepton number $L_{e0} = 4.8 \times 10^{-3}$.  The sterile neutrino properties are just below the current x-ray constraints, while the initial lepton number is much less than the best constraints on the lepton numbers in the early universe.  We display only the energy spectrum because the other three panels do not provide much physical insight in this case.

For this case our calculations give $\epsilon_{\rm max} \approx 200$, which means that the resonance sweeps through the entire $\nu_e$ distribution.  As a result, the high-$\epsilon$ peak cannot be seen in Fig.\ \ref{fig:inplay}.  The mid-$\epsilon$ peak is at $\epsilon \approx 60$, so that is also missing.  We are left with the low-$\epsilon$ peak and a long tail at higher $\epsilon$ that eventually disappears because of the scarcity of neutrinos in the $\nu_e$ distribution at these scaled momenta.

For this particular case the resonance cessation event occurs at $T \approx 100 \,{\rm MeV}$, but as discussed above, this corresponds to $\epsilon \approx 60$.  The low-temperature resonance for more relevant values of $\epsilon$ ({\it i.e.}, $\epsilon \sim 3$) occurs at temperatures $\sim 170 \,{\rm MeV}$.  Since the resonances persist to such low temperatures, there is very little nonresonant production of sterile neutrinos.  As a result, the relic sterile neutrino density produced is relatively insensitive to the total scattering rate.  However, this temperature is of interest in the early universe as it is the epoch of the QCD transition.  The nature of this transition affects the specifics of the sterile neutrino production \cite{af02} but in broad brush should serve to enhance the final sterile neutrino closure fraction.

Also plotted on Fig.\ \ref{fig:inplay} is a Fermi-Dirac spectrum whose normalization is chosen to have an equivalent sterile neutrino density.  The peak in the sterile neutrino spectrum is at $\epsilon \approx 1$, while the Fermi-Dirac spectrum peaks at $\epsilon \approx 2.2$.  We see that a significant fraction of the total sterile neutrino density has smaller values of $\epsilon$ than in its Fermi-Dirac counterpart.  Nonetheless, the high-$\epsilon$ tail of the distribution yields an average value of $\epsilon$, $\langle \epsilon \rangle \approx 2$, compared to an average $\epsilon$ of 3.15 for a Fermi-Dirac spectrum.  As a result, the sterile neutrino population produced in our calculations is ``colder'' by a factor of about 1.6 than an equivalent population with a Fermi-Dirac spectrum.  This is qualitatively consistent with the findings in Ref.\ \cite{sf99}, but the sterile neutrino population here is not as cold.

Our numerical calculations show that the sterile neutrino rest mass, vacuum mixing angle, and initial lepton number parameters which produce the correct dark matter closure fraction, but which evade X-ray bounds, produce relic $\nu_s$ energy spectra which are qualitatively similar to that in Fig.\ \ref{fig:inplay}.  Since this generic spectrum is skewed relative to a Fermi-Dirac form toward lower values of $\epsilon$, viable dark matter sterile neutrinos which are produced via lepton number enhancement should be ``colder'' than their rest masses would suggest.

\begin{figure}
\includegraphics[width = 2in, angle = 270]{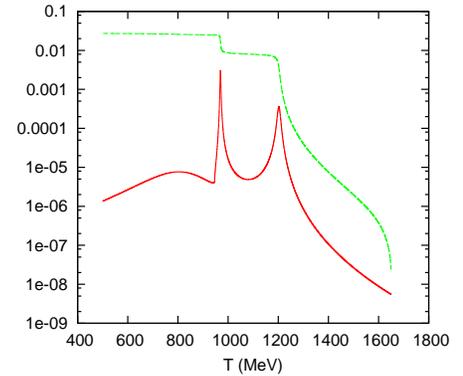}
\caption{\label{fig:zenocomp} The solid curve is the sterile neutrino conversion rate per unit temperature interval, $\vert d f_s / d T \vert$ for $\epsilon = 1$ in our calculations with $m_s = 64 \,{\rm keV}$, $\sin^2 2 \theta = 10^{-10}$, and $L_{e0} = 1.1 \times 10^{-3}$.  The dashed curve shows the fractional difference between the rate inferred from a solution of the quantum kinetic equations to the rate implied by the quantum Zeno ansatz.}
\end{figure}

Finally, in Fig.\ \ref{fig:zenocomp} we perform a consistency check to determine the validity of the quantum Zeno ansatz that we have used throughout our calculations.  To do this, we use the sterile neutrino parameters and initial lepton numbers employed in the calculations for Fig.\ \ref{fig:ex} ($m_s = 64 \,{\rm keV}$, $\sin^2 2 \theta = 10^{-10}$, and $L_{e0} = 1.1 \times 10^{-3}$), and also use the smaller total scattering rate discussed above.  We solve the quantum kinetic equations presented in Sec.\ \ref{sec:qke} for $\epsilon = 1$, using the lepton number evolution from our calculations.  

Plotted in Fig.\ \ref{fig:zenocomp} are the instantaneous $\nu_e \rightarrow \nu_s$ conversion rate per unit temperature interval, $\vert d f_s / d T \vert$ (solid curve), and the fractional difference between the rate inferred from our solution to the quantum kinetic equations and the rate given by the quantum Zeno ansatz (dashed curve).  Note that the actual conversion rate is always slightly larger than the rate from the quantum Zeno effect, but in general the two agree within a few percent.  Note also that the differences change by an order of magnitude at the resonances.  

Since the actual rate is possibly slightly higher than the rate we used with the quantum Zeno ansatz, we probably under-produce sterile neutrinos in our calculations.  The interval between the two resonances is an important one in determining the final sterile neutrino production.  If corrections on the order of a percent occurred for all our values of $\epsilon$, this could increase the temperature of the resonance cessation event and, in turn, increase the ultimate relic sterile neutrino density, possibly in a nonlinear fashion.  Our numerical calculations show that a 1 \% increase in the conversion rate gives a 0.3 \% increase in relic sterile neutrino density.  We conclude that a small change in the conversion rate does not lead to a significant change in the relic sterile neutrino density.

\section{Conclusions}
\label{sec:conclusions}

One option for producing viable sterile neutrino dark matter is lepton number enhancement of scattering-induced decoherence.  The calculations presented here address the basic physics of this production process.  A key conclusion of our work is that the interplay of lepton number depletion and resonant sterile neutrino production leads to a set of generic peak features in the relic sterile neutrino energy spectrum.  We have studied how these peaks depend on the assumed initial lepton number, the magnitude of the assumed active-sterile coupling ({\it i.e.}, the vacuum mixing angle), and the sterile neutrino rest mass.  We conclude that relic sterile neutrinos with parameters that evade x-ray constraints will have a characteristic single-peak energy spectrum.  This spectrum will be ``cold'' compared to a Fermi-Dirac form spectrum with the same integrated relic number density.  This could relax to some extent Lyman-alpha forest constraints on sterile neutrino rest mass \cite{b08}.

We have compared the widely-used quantum Zeno ansatz with a full quantum kinetic equation treatment of scattering-induced decoherence production of sterile neutrinos in the early universe.  Our conclusion is that the quantum Zeno approximation is adequate, at least during the epoch in the early universe when $\sim \,{\rm keV}$ mass sterile neutrinos are produced.

Finally, we should note the limitations of this study and point out the topics that need further detailed examination.  The calculations presented here are crafted to study the physical effects which determine relic sterile neutrino energy spectra.  They only treat schematically the microphysics of scattering degrees of freedom and finite temperature in-medium effects.  We have attempted to gauge the sensitivity of final relic $\nu_s$ densities and energy spectra to these issues by simply varying the prescription  for active neutrino scattering rates.  Though we find that our generic spectral features are relatively insensitive to changes in these rates, our calculations nevertheless cannot serve to predict relic densities accurately.

At least three aspects of our calculations for sterile neutrino production and lepton number depletion are simplistic:  (1) we have allowed only a single channel, $\nu_e \rightarrow \nu_s$, for sterile neutrino production; (2) we have taken initial, preexisting lepton numbers in all active neutrino species to be the same; and (3) we have not allowed mixing between active neutrino flavors. As for the last point, the study in Ref.\ \cite{abfw} suggests that dynamically including active-active mixing should be roughly similar to having a larger initial value of $L_{e0}$ in our model. The idea being that as $\nu_e$ are converted to $\nu_s$, some $\nu_\mu$ and $\nu_\tau$ neutrinos convert into $\nu_e$. In essence then, the net lepton number in the mu and tau neutrino seas serves as a reservoir that ``feeds'' the $\nu_e$ sea as sterile neutrinos are produced.  In fact, the measured active-active mixing parameters may suggest mixing between the flavors is efficient and that equality of lepton numbers is a plausible assumption \cite{abb02}. A larger reservoir of active neutrinos feeding into $\nu_e \rightarrow \nu_s$ would not change the {\it qualitative} aspects of our conclusions on the interplay of lepton number depletion and neutrino asymmetry-enhanced sterile neutrino production in the early universe.

In contrast, if there were more than one kind of sterile neutrino, or if there were multiple channels of active-sterile conversion ({\it e.g.,} $\nu_{\mu,\tau } \rightarrow \nu_s$ in addition to $\nu_e \rightarrow \nu_s$), then both lepton number depletion and sterile neutrino production histories might be different than those calculated here. In fact, a key conclusion of our work is that these histories, as well as the relic sterile neutrino energy spectrum that goes with them, are a product of the potentially complicated coupling between the sterile neutrino production channels and the way lepton numbers evolve. This is a rich problem which deserves further consideration.

\begin{acknowledgments}
This work was supported in part by NSF Grant PHY-04-00359 at UCSD.  C.T.K. would like to thank J.\ Kao for useful discussions on the computational aspects of this work and the ARCS Foundation, Inc. for financial support.  We would like to thank A.\ Kusenko for valuable conversations.
\end{acknowledgments}

\bibliography{dm}

\end{document}